\documentclass[preprint,12pt]{elsarticle}

\usepackage{url,amssymb,amsmath,amsthm,xspace,wrapfig,pifont,bbding,nicefrac}

\journal{Information Processing Letters}

\usepackage[T1]{fontenc}

\newtheorem{theorem}{Theorem}[section]

\newtheorem{remark}[theorem]{Remark}

\newenvironment{frcseries}{\fontfamily{emerald}\selectfont}{}
\newcommand{\textfrc}[1]{{\frcseries#1}}

\newcommand{\comment}[1]{}

\newcommand{\poly}{\mathrm{poly}}

\newcommand{\M}{\mathsf{M}}

\newcommand{\D}{\mathsf{D}}

\newcommand{\NP}{\mathsf{NP}}

\newcommand{\opt}{\mathsf{OPT}}

\newcommand{\cS}{\mathbf{S}}
\newcommand{\cC}{\mathcal{C}}

\newcommand{\eps}{\varepsilon}

\renewcommand{\epsilon}{\varepsilon}
\newcommand{\IE}{{\em i.e.}\xspace}

\newcommand{\rank}{{{\mathrm{rank}}}}

\biboptions{sort&compress}

\begin{document}

\begin{frontmatter}

\title{On a Connection Between Small Set Expansions and Modularity Clustering}

\author{Bhaskar DasGupta\fnref{f1}\corref{f2}}
\ead{dasgupta@cs.uic.edu}
\ead[url]{www.cs.uic.edu/~dasgupta}
\address{Department of Computer Science, University of Illinois at Chicago, Chicago, IL 60607}
\fntext[f1]{Partially supported by NSF grants IIS-1160995.}
\cortext[f2]{Corresponding author.}

\author{Devendra Desai}
\ead{devdesai@cs.rutgers.edu} 
\ead[url]{paul.rutgers.edu/~devdesai}
\address{Department of Computer Science, Rutgers University, Piscataway, NJ 08854}

\begin{abstract}
In this paper we explore a connection between two seemingly different problems from two different domains: 
the {\em small-set expansion} problem studied in unique games conjecture, and a popular community finding approach
for social networks known as the {\em modularity clustering} approach. We show that a sub-exponential time algorithm for the small-set expansion problem
leads to a sub-exponential time constant factor approximation for some hard input instances of the modularity clustering problem. 
\end{abstract}

\begin{keyword}
Small-set expansion \sep modularity clustering \sep social network

\MSC[2010] 68Q25 \sep 68W25
\end{keyword}

\end{frontmatter}

\section{Introduction and Definitions}

All graphs considered in this note are {\em undirected} and {\em unweighted}\footnote{Our result can be extended for 
the more general case of directed weighted graphs by using the correspondence of these versions with unweighted undirected graphs
as outlined in~\cite[Section $5.1$]{our-arxiv}.}. Let $G=(V,E)$ denote the given input graph with $n=|V|$ nodes and $m=|E|$ edges, 
let $d_v$ denote the degree of a node $v\in V$, and 
let $A(G)=\big[a_{u,v}(G)\big]$ denote the adjacency matrix of $G$, \IE, 
$a_{u,v}(G)=
\left\{
\begin{array}{ll}
1, & \mbox{if $\{u,v\}\in E$} \\ 
0, & \mbox{otherwise.}
\end{array}
\right.$
Since our result spans over two distinct research areas, we summarize the relevant definitions from both research fields~\cite{N06,ABS10} below
for convenience.

\vspace*{0.1in}
\noindent
{\bf (\textfrc{a})}
By a ``set of ($k$) {\em communities}'' we mean a partition of the set of nodes $V$ into ($k$) non-empty parts.

\vspace*{0.1in}
\noindent
{\bf (\textfrc{b})}
If $G$ is $d$-regular for some given $d$, then its symmetric {\em stochastic walk} matrix is 
denoted by $\widehat{A}(G)$, and is defined as the $n\times n$ real symmetric matrix 
$\widehat{A}(G)=\left[\frac{a_{u,v}(G)}{d}\right]$.

\vspace*{0.1in}
\noindent
{\bf (\textfrc{c})}
For a real number $\tau\in[\,0,1)$, the {\em $\tau$-threshold rank} of $G$, denoted by $\rank_\tau(G)$, is the number of
eigenvalues $\lambda$ of $\widehat{A}(G)$ satisfying $|\lambda|>\tau$.

\vspace*{0.1in}
\noindent
{\bf (\textfrc{d})}
For a subset $\emptyset\subset S\subset V$ of nodes, the following quantities are defined:
\begin{itemize}
\item
The (normalized) {\em measure} of $S$ is $\mu(S)=\frac{|S|}{n}$.

\item
The (normalized) {\em expansion} of $S$ is 
$
\Phi(S)=\dfrac{\big|\,\big\{ \, \{u,v\}\,|\,u\in S,\,v\not\in S,\,\{u,v\}\in E \big\} \, \big|}{\sum\limits_{v\in S} d_v}
$
\label{pp}
\item 
The (normalized) {\em density} of $S$ is $\D(S)=1-\Phi(S)$. 
\item
The {\em modularity} value of $S$ is 
$
\M(S)=
\frac{1}{2m} \left(
\sum\limits_{\,u,v\in S} 
\left(
a_{u,v}-\frac{d_u d_v}{2m}
\right)
\right) 
$
\end{itemize}

\vspace*{0.1in}
\noindent
{\bf (\textfrc{e})}
The modularity of a set of communities $\cS$ is $\M(\cS)=\sum_{S\in\,\cS}\M(S)$.

\vspace*{0.1in}
\noindent
{\bf (\textfrc{f})}
The goal of the {\em modularity $k$-clustering} problem on an input graph $G$ is to find a set of at most $k$ 
communities $\cS$ that {\em maximizes} $\M(\cS)$. 
Let $\displaystyle\opt_k(G)=\max_{\cS\ \mathrm{is\,\, a\,\, set\,\, of\,\,at\,\,most\,\,k\,\,communities}} \big\{\,\M(\cS)\,\big\}$ denote the optimal modularity value
for a modularity $k$-clustering; it is easy to verify that $0\leq\opt_k(G)<1$.

\vspace*{0.1in}
\noindent
{\bf (\textfrc{g})}
The goal of the {\em modularity clustering} problem on $G$ is to find a set of (unspecified number of) 
communities $\cS$ that {\em maximizes} $\M(\cS)$. 
Let $\displaystyle\opt(G)$ denote the optimal modularity value for a modularity clustering; obviously, $\opt(G)=\opt_n(G)$.

\vspace*{0.1in}
\noindent
{\bf (\textfrc{h})}
$\exp(\xi)$ denotes $2^{\,c\xi}$ for some constant $c>0$ that is independent of $\xi$.

\vspace*{0.1in}
The modularity clustering problems as described above is {\em extremely popular} in practice in their applications to 
biological networks~\cite{m1,m2} as well as to social networks~\cite{LN08,NG04,N06}.
For relevant computational complexity results for modularity maximization, see~\cite{our-arxiv,BDGGHNW07-1}.
The following results from~\cite{our-arxiv} demonstrate the computational hardness of $\opt_2(G)$ and 
$\opt(G)$ even if $G$ is a regular graph.

\begin{theorem}~{\rm$\!\!\!\!$\cite{our-arxiv}}\label{aa}~\\

\vspace*{-0.1in}
\noindent
{\bf (a)}
For every constant $d\geq 9$, there exists a collection of $d$-regular graphs $G$ of $n$ nodes such it is
$\NP$-hard to decide if $\opt_2(G)\geq \frac{1}{2}-\frac{2c}{dn}$ or if $\opt_2(G)\leq \frac{1}{2}-\frac{2c+2}{dn}$
for some positive $c=O(\sqrt{n}\,)$.

\vspace*{0.1in}
\noindent
{\bf (b)}
There exists a collection of $(n-3)$-regular graphs $G$ of $n$ nodes such it is
$\NP$-hard to decide if $\opt(G)>\frac{0.9388}{n-4}$ or if $\opt(G)<\frac{0.9382}{n-4}$.
\end{theorem}

\section{Our Result}

\begin{theorem}\label{main-thm}
Let $G$ be a $d$-regular graph. Then, 
for some constant $0<\eps<\nicefrac{1}{2}$, there is an algorithm $\mathcal{A}_\eps$ with the
following properties:
\begin{itemize}
\item
$\mathcal{A}_\eps$ runs in sub-exponential time, \IE,
in time $\exp(\delta\,n)$ for some constant $0<\delta=\delta(\eps)<1$ that depends on $\eps$ only.

\item
$\mathcal{A}_\eps$ correctly distinguishes instances $G$ of modularity clustering with $\opt(G)\geq 1-\eps$ from instances $G$ with $\opt(G)\leq\eps$. 
\end{itemize}
$($Note that we make no claim if $\eps<\opt(G)<1-\eps$.$)$
\end{theorem}

\begin{remark}[\bf usability of the approximation algorithm in Theorem~\ref{main-thm}]
We prove Theorem~\ref{main-thm} for $\eps=10^{-6}$. 
It is natural to ask if there are in fact infinite families of $d$-regular graphs $G$ that satisfy 
$\opt(G)\geq 1-10^{-6}$ or $\opt(G)\leq 10^{-6}$. 
The answer is affirmative, and we provide below examples of infinite families of such graphs.

\begin{quote}
\vspace*{0.1in}
\noindent
$\mathbf{\opt(G)\geq 1-10^{-6}}$: 
Consider, for example, the following explcit bound was demonstrated in~{\rm\cite[Corollary 6.4]{BDGGHNW07-1}}:
\begin{quote}
if $G$ is an union of $k$ disjoint cliques each with $\frac{n}{k}>3$ nodes then $\opt(G)=1-\frac{1}{k}$.
\end{quote}
\noindent
Based on this and other known results on modularity clustering, examples of families of regular graphs $G$ for which $\opt(G) \geq 1-10^{-6}$ include:
\begin{description}
\item[(1)]
$G$ is an union of $k$ disjoint cliques each with $\frac{n}{k}>3$ nodes for any $k>10^6$.

\item[(2)]
$G$ is obtained by a local modification from the graph in {\bf (1)} such as: 
\begin{itemize}
\item 
Start with an union of $k$ disjoint cliques $\cC_1,\cC_2,\dots,\cC_k$ each with $\frac{n}{k}>3$ nodes for any $k$ sufficiently large with respect to $10^6$ ($k\geq 10^7$ suffices).

\item
Remove an arbitrary edge $\{u_i,v_i\}$ from each clique $\cC_i$. Let $U=\cup_{i=1}^k \left\{ u_i \right\}$ and 
and $V=\cup_{i=1}^k \left\{ v_i \right\}$.

\item
Add to $G$ the edges corresponding to any perfect matching in the complete bipartite graph with node sets $U$ and $V$.
\end{itemize}
\end{description}

\vspace*{0.1in}
\noindent
$\mathbf{\opt(G)\leq 10^{-6}}$: 
Theorem~\ref{aa}~{\rm\cite{our-arxiv}} involves infinitely many graphs of $n>4+0.9388\times 10^6$ nodes satisfying $\opt(G) < \frac{0.9388}{n-4}<10^{-6}$
(these graphs are edge complements of appropriate families of $3$-regular graphs used in~{\rm\cite{CC06}}).
\end{quote}
\end{remark}

\begin{proof}[Proof of Theorem~\ref{main-thm}]$\!\!\!$\footnote{We have made no significant attempts to optimize the constants in Theorem~\ref{main-thm}.}
Set $\eps=10^{-6}$. We assume that $G$ is $d$-regular, and either $\opt(G)\geq 1-10^{-6}$ or $\opt(G)\leq 10^{-6}$.

\subsection*{Preliminary Algebraic Simplification}

Let $\cS=\big\{S_1,S_2,\dots,S_k\big\}$ be a set of communities of $G$.
The objective function $\M(\cS)$ can be equivalently expressed as follows via simple algebraic manipulation~\cite{BDGGHNW07-1,LN08,NG04,N06}.
Let $m_i$ denote the number of edges whose both endpoints are in $S_i$, 
$m_{ij}$ denote the number of edges one of whose endpoints is in $S_i$ and the other in $S_j$ 
and $D_i = \sum\limits_{v\in S_i} d_v$ denote the sum of degrees of nodes in $S_i$.
Then, 
$\M(\cS)=
\sum\limits_{S_i\in\,\cS}\, 
\left(
\frac{m_i}{m} - \left( \frac{D_i}{2m} \right)^2 
\right)$.

We will provide an approximation for $\opt_2(G)$ and then use the result that $\opt_2(G)\geq\frac{\opt(G)}{2}$ proved in~\cite{our-arxiv}.
Note that if if $\opt(G)\leq 10^{-6}$ then obviously $\opt_2(G)\leq 10^{-6}$, whereas if $\opt(G)\geq 1-10^{-6}$ then 
$\opt_2(G)\geq \frac{1}{2}\,-\,\frac{10^{-6}}{2}$. 
Consider a partition $\cS$ of $V$ into exactly two sets, say $S$ and $\overline{\rule[8pt]{0pt}{2pt}S}=V\setminus S$ 
with $0<\mu(S)\leq \nicefrac{1}{2}$. By Lemma~2.2 of~\cite{our-arxiv}, $\M(S)=\M(\overline{\rule[8pt]{0pt}{2pt}S})$ and thus 
\[
\begin{array}{lll}
\M(\cS) = 2\times \left( \dfrac{m_1}{m} - \left( \dfrac{|S|}{n} \right)^{\!\!2\,} \right)
&
=
&
2\times \left( \dfrac{\frac{1}{2}\,\,\D(S)\,d\,|S|}{\frac{1}{2}\,\,d\,n} - \mu(S)^2\right)
\\
\\
& = & 
2\times \big(\, \D(S)\,\mu(S) - \mu(S)^2 \, \big)
\end{array}
\]
Thus, letting $\D=\D(S),\mu=\mu(S)$ and $\Phi=\Phi(S)$, we have $\Phi=1-\D$ as per our notations used in page~\pageref{pp} and
the goal of modularity $2$-clustering is to maximize the following function $f$
over all possible valid choices of $\D$ and $\mu$: 
\[
f(\mu,\D) 
= 
2\times \left(\mu\,\D - \mu^2\right)
=
2\times \left(\,\mu(1-\Phi) - \mu^2 \,\right)
\]
Let $\cS^\star=\{\,S^\star,\overline{\rule[8pt]{0pt}{2pt}S^\star}\,\}$ be an optimal solution for modularity $2$-clustering of $G$,
with $\D=\D^\star,\mu=\mu^\star,\Phi=\Phi^\star$ (and thus $\opt_2(G)=f(\mu^\star,\D^\star)\,$). Obviously,
\begin{gather*}
\left|\mu^\star-\frac{\D^\star}{2}\right|\,<\,\frac{\D^\star}{2} \\
f\left(\frac{\D^\ast}{2}+\delta,\D^\ast\right)=f\left(\frac{\D^\ast}{2}-\delta,\D^\ast\right) \mbox{  for any positive  } \delta>0
\end{gather*}
Note that we need to show that, if 
$\opt_2(G)=f(\mu^\star,\D^\star)>\frac{1}{2}-\frac{10^{-6}}{2}$, then there is an algorithm $\mathcal{A}_{\eps}$ 
as described in Theorem~\ref{main-thm} that outputs a valid choice of $\mu$ and $\D$, say 
$\mu'$ and $\D'$, such that $f(\mu',\D')>10^{-6}$.

\subsection*{Guessing $\D^\star$} 

Note that there are at most $\mathrm{O}\left(d\,n^2\right)$ choices for $\D^\star$ since $\D^\star$ is of the form 
$\nicefrac{i}{(j\,d)}$ for $j\in\left\{1,2,\dots,\nicefrac{n}{2}\,\right\}$ and $i\in \big\{1,2,\dots,j\,d\,\big\}$. 
In the sequel, we will run our algorithm for each choice of $\D^\star$ and take the best of these 
solutions. Thus, it will suffice to prove our approximation bound assuming we have guessed $\D^\star$ exactly.

In the remainder of the proof, we will make use of results for small-set expansion from~\cite{ABS10}. 
The description is self-contained, and the reader {\em will not need any prior knowledge of expansion properties of graphs}.
Remember that we assume that $f(\mu^\star,\D^\star)>\frac{1}{2}\,-\,\frac{10^{-6}}{2}$ and thus $\mu^\star>\frac{1}{2}\,-\,\frac{10^{-3}}{2}$
since otherwise
\[
\begin{array}{r l}
&
\mu^\star\leq \frac{1}{2}\,-\,\frac{10^{-3}}{2}
\\
\Rightarrow 
&
\mu^\star=\frac{1}{2}\,-\,\frac{10^{-3}}{2} -\xi 
\hspace*{0.3in}
\mbox{[ for some $\xi\geq 0$ ]}
\\
\Rightarrow 
&
f(\mu^\star,\D^\star) 
= 
2\times \left(\mu^\star\,\D^\star - \left(\mu^\star\right)^2\right)
\\
&
\hspace*{0.7in}
\leq
2\times \left(\mu^\star - \left(\mu^\star\right)^2\right)
\hspace*{0.3in}
\mbox{[ since $0\leq \D^\star\leq 1$ ]}
\\
&
\hspace*{0.7in}
=
2\times \mu^\star\times \left(1 - \mu^\star\right)
\\
&
\hspace*{0.7in}
=
2 \times 
\left( \frac{1}{2}\,-\,\frac{10^{-3}}{2} -\xi \right)
\times
\left( \frac{1}{2}\,+\,\frac{10^{-3}}{2} +\xi \right)
\\
&
\hspace*{0.7in}
=
2 \times 
\left( \frac{1}{4}\,-\,\left(\frac{10^{-3}}{2} +\xi\right)^2 \, \right)
\\
&
\hspace*{0.7in}
<
\frac{1}{2} - \frac{10^{-6}}{2}
\end{array}
\]
which contradicts
$f(\mu^\star,\D^\star)>\frac{1}{2}\,-\,\frac{10^{-6}}{2}$.
Similarly, we also get:
\[
\D^\star  = \dfrac{f(\mu^\star,\D^\star)}  { 2\,\mu^\star  } + \mu^\star
> 
\left(\dfrac{1-10^{-6}}{4} \right)\,\frac{1}{\mu^\star} +\mu^\star
\]
Consider the function 
$g(\mu)=\frac{a}{\mu}+\mu$ where $a=\frac{1-10^{-6}}{4}$.
Since $\mu>0$, 
$\frac{ \mathrm{d}^2 \, g(\mu)}{\mathrm{d} \, \mu^2}=\frac{2a}{\mu^3}>0$ and thus the minimum of $g(\mu)$ is attained at $\mu=b$ 
that satisfies $\frac{ \mathrm{d} \, g(b)}{\mathrm{d} \, b}=\,-\,\frac{a}{b^2}+1=0$, giving $b=\sqrt{a}$.
Thus, we have 
\[
\D^\star
>
\left( \frac{1-10^{-6}}{4} \right)\,
\left(\frac{1}{ \sqrt{ \frac{1-10^{-6}}{4} } } \right)
\,+\,
\sqrt{ \frac{1-10^{-6}}{4} }
=\sqrt{1-10^{-6}}>1-10^{-6} 
\]
which implies $\Phi^\star=1-\D^\star<10^{-6}$.

\vspace*{0.1in}
\noindent
{\bf Case I: $G$ has a small threshold rank, \IE, $\pmb{\rank_{\,1-10^{-6}}(G)<n^{10^{-1}}}$} 

\vspace*{0.1in}
The following result, restated below under the assumption of this case in our terminologies after instantiation of 
parameters with specific values and trivial algebraic simplification,
was proved by Arora, Barak and Steurer in~\cite{ABS10}
in the bigger context of obtaining sub-exponential algorithms for unique games in {\sf PCP} theory.

\begin{theorem}{\rm\cite{ABS10}}\footnote{Instantiate Theorem~2.2 in~\cite{ABS10} with $\eta=10^{-4}$ and $\eps=10^{-6}$.}\label{thm11}
There exists a $\left(\exp\left(n^{10^{-1}}\right)\poly(n)\right)$-time algorithm that
outputs a subset $\emptyset\subset S\subset V$ such that 
$0.92\,|\,S^\star|\leq |\,S|\leq 1.08\,|\,S^\star|$, and 
$\Phi(S)\leq\Phi(S^\star)+0.08$.  
\end{theorem}

We run the algorithm in Theorem~\ref{thm11}, and return $\big\{\,S,\overline{\rule[8pt]{0pt}{2pt}S}\,\big\}$ as our solution.
Note that:
\begin{gather*}
\Phi(S)\leq\Phi^\star+0.08<0.080001\, \Longrightarrow\, \D(S)>1-0.080001=0.919999 \\
0.92\mu^\star\leq \mu(S)\leq 1.08\mu^\star \,\Longrightarrow\, 0.4599\leq\mu(S)\leq 0.54
\end{gather*}
and thus 
\[
\begin{array}{lll}
f(\mu(S),\D(S)) & = & 2\times \mu(S)\times \big(\, \D(S) - \mu(S)\,\big)
\\
& > & 2\times 0.4599 \times \big(0.919999-0.54\big) > 10^{-6}
\end{array}
\]

\vspace*{0.1in}
\noindent
{\bf Case II: Remaining Case, \IE, $\pmb{\rank_{\,1-10^{-6}}(G)\geq n^{10^{-1}}}$} 

\vspace*{0.1in}
The following result, restated below in our terminologies after instantiation of 
parameters with specific values, was again proved in~\cite{ABS10}.

\begin{theorem}{\rm\cite{ABS10}}\footnote{Instantiate Theorem~2.3 in~\cite{ABS10} with $\eta=10^{-4}$ and $\gamma=10^{-1}$.}\label{thm12}
Let $H$ be a regular graph of $r$ nodes with $\rank_{\,1-10^{-5}}(H)\geq r^{{10}^{-1}}$.
Then, there is an algorithm that 
\begin{itemize}
\item
runs in $\poly(r)$ time, and 

\item
finds a subset $S$ of nodes of $H$ with $|\,S|\leq r^{1-10^{-3}}$ and $\Phi(S)\leq 10^{-2}$.
\end{itemize}
\end{theorem}

Our strategy is to use the algorithm in Theorem~\ref{thm12} 
{\em repeatedly}\footnote{\cite{ABS10} points
out how to ``re-regularize'' the remaining graph each time a set of nodes have been extracted by adding appropriate number
of self-loops of weight $\nicefrac{1}{2}$.}
to extract ``high-rank parts'' from $G$.
Namely, we compute in polynomial time an ordered partition of nodes $\big(\,T_1,T_2,\dots,T_k,V\setminus\cup_{i=1}^k T_i\,\big)$ such that 
each $T_i$ is obtained by using the algorithm in Theorem~\ref{thm12} on graph $G_i$ induced by the set of nodes 
$V\setminus\cup_{j=1}^{i-1} T_i$, and the last (possibly empty) graph $G''$ induced by the set of nodes 
$V''=V\setminus\cup_{i=1}^k T_i$ satisfy $\rank_{\,1-10^{-6}}(G'')<|V''|^{10^{-1}}$. Let $G'$ be the graph induced 
by the set of nodes $V'=\cup_{i=1}^k T_i$.  

\vspace*{0.1in}
\noindent
{\bf Case II(a)}
$\displaystyle\pmb{\big|\,S^\star\cap V''\big|\geq\nicefrac{|S^\ast|}{2}}$.

\vspace*{0.1in}
Let $S_1^\ast$ be the set containing an arbitrary $\nicefrac{|S^\ast|}{2}$ elements from the set $S^\star\cap V''$.
Note that $\mu\left(S_1^\ast\right)=\nicefrac{\mu^\ast}{2}$ and $\Phi\left(S_1^\ast\right)\leq 2\,\Phi^\ast$.
We now use Theorem~\ref{thm11} on the graph $G''$ with $|S^\ast|$ replaced by $\nicefrac{|S^\ast|}{2}$ to output a set $S\subseteq V''$ 
of nodes such that 
\begin{gather*}
\Phi(S)\leq 2\,\Phi^\star+0.08<0.080002\, \Longrightarrow\, \D(S)>1-0.080002=0.919998 \\
0.46\mu^\star\leq \mu(S)\leq 0.54\mu^\star \,\Longrightarrow\, 0.229<\mu(S)\leq 0.27
\end{gather*}
and thus 
\[
\begin{array}{lll}
f(\mu(S),\D(S)) & = & 2\times \mu(S)\times \big( \, \D(S) - \mu(S)\, \big)
\\
 & > & 2\times 0.229 \times (0.919998-0.27) > 10^{-6}
\end{array}
\]

\vspace*{0.1in}
\noindent
{\bf Case II(b)} 
$\displaystyle\pmb{\big|\,S^\star\cap V''\big|<\nicefrac{|S^\ast|}{2}}$.

\vspace*{0.1in}
Since $|S^\ast|\geq \left(\frac{1}{2}\,-\,\frac{10^{-3}}{2}\right)n$ and 
$|T_j|\leq \left|\,V\setminus \cup_{\ell\,=1}^{j-1}T_{\ell} \,\right|^{1-10^{-3}}<n^{1-10^{-3}}$ for any $j$,
there exists an index $i$ such that 
$\frac{|S^\ast|}{2}-n^{1-10^{-3}}<\left|\cup_{j=1}^i T_j\right|<\frac{|S^\ast|}{2}+n^{1-10^{-3}}$.
Notice that the graph induced by the set of nodes 
$S=\cup_{j=1}^i T_j$ satisfy $\Phi(S)\leq 10^{-2}$ and, 
since $\left(\frac{1}{2}-\frac{10^{-3}}{2}\right)n \leq |S^\ast|\leq n$, 
we have 
\begin{gather*}
\frac{|S^\ast|}{2}-n^{1-10^{-3}} < |S| = \left|\cup_{j=1}^i T_j\right| < \frac{|S^\ast|}{2}+n^{1-10^{-3}}
\Longrightarrow
0.24<\mu(S)<0.51
\end{gather*}
and thus, 
$
\begin{array}{lll}
\\
f(\mu(S),\D(S))
& 
=
& 
2\times \mu(S)\times \left( \,\D(S) - \mu(S)\,\right)
\\
& 
>
& 
2\times 0.24 \times (0.99-0.51)
>
10^{-6}
\end{array}
$

\end{proof}

\section*{Further Research}

An interesting open question is whether it is possible to prove the converse of
Theorem~\ref{main-thm}, \IE, 
can we use a sub-exponential approximation algorithm for modularity 
maximization to design a sub-exponential algorithm for small-set expansion problems ?
If possible, this may lead to an alternate interpretation of unique games 
via communities in social networks.

\vspace*{0.1in}
\noindent
{\bf Acknowledgements}
We thank Mario Szegedy for pointing out reference~\cite{ABS10}.

\end{document}